\pdfoutput=1
\documentclass[lettersize,journal]{IEEEtran}
\usepackage{amsmath,amsfonts}
\usepackage{array}
\usepackage{multirow}
\usepackage{textcomp}
\usepackage{stfloats}
\usepackage{url}
\usepackage{graphicx}
\usepackage{cite}
\usepackage{orcidlink}
\hypersetup{hidelinks}
\graphicspath{{figures/}}
\begin{document}

\title{Denoising Low-Latitude Ionograms Without Losing Spread-F Echoes}

\author{Mert~Can~Turkmen~\orcidlink{0009-0003-6195-1715},
Septi~Perwitasari~\orcidlink{0000-0002-5108-0077},
Eng~Leong~Tan~\orcidlink{0000-0003-3003-6007},
Pornchai~Supnithi,
Le~Truong~Thanh,
and~Yee~Hui~Lee~\orcidlink{0000-0001-6452-9606}%
\thanks{M. C. Turkmen, E. L. Tan, and Y. H. Lee are with the School of Electrical and Electronic Engineering, Nanyang Technological University (NTU), Singapore 639798 (e-mail: mertcan001@e.ntu.edu.sg; eeltan@ntu.edu.sg; eyhlee@ntu.edu.sg).}%
\thanks{S. Perwitasari is with the National Institute of Information and Communications Technology (NICT), Koganei, Tokyo, Japan (e-mail: septi.perwitasari@nict.go.jp).}%
\thanks{P. Supnithi is with the School of Engineering, King Mongkut's Institute of Technology Ladkrabang (KMITL), Bangkok, Thailand (e-mail: pornchai.su@kmitl.ac.th).}%
\thanks{L. T. Thanh is with the Institute of Earth Sciences, Vietnam Academy of Science and Technology (IES-VAST), Hanoi, Vietnam (e-mail: igpthanh@gmail.com).}%
\thanks{Corresponding author: M. C. Turkmen (e-mail: mertcan001@e.ntu.edu.sg).}}

\markboth{Preprint}%
{Turkmen \MakeLowercase{\textit{et al.}}: A Density-Based Clustering Denoiser for Low-Latitude Ionogram Images}

\maketitle

\begin{abstract}
Ionograms from the SEALION FMCW ionosondes carry vertical interference bands and speckle noise. After sunset, echo traces can spread into a diffuse cloud that looks much like this speckle. Published cleaning methods focus only on line-like traces. This letter presents a denoising method that keeps thin traces, multiple hops and spread-F echoes. It combines a column-median background filter, DBSCAN on tilted pixel coordinates and rule-based cluster filters. On 100 randomly sampled frames graded by eye, no echo was lost outside heavy interference, and 19 of 21 spread-F echoes were kept whole. The code and examples of the cleaned ionograms are released at \url{https://github.com/Mert-chan/ionogram-denoiser}.
\end{abstract}


\section{Introduction}
\IEEEPARstart{A}{n} ionosonde is a high-frequency radar that sweeps a range of frequencies to sound the vertical profile of the ionosphere. The resulting figure, with virtual height on the y-axis and frequency on the x-axis, is called an ionogram. Ionograms are one of the primary tools used to study the ionosphere, particularly the F region, which is the layer where most of the electron density is concentrated. In Southeast Asia, the SEALION observation network \cite{maruyama2007} operates ionosondes near the magnetic equator to observe post-sunset irregularities in the F region, which cause scintillation of satellite signals \cite{kintner2007}.

The SEALION ionosondes are portable frequency-modulated continuous-wave (FMCW) sounders whose receivers have no tuning at the input stage, apart from a 2~MHz high-pass filter. Thus, broadcast and communication signals across the sweep enter the receiver together with the echo \cite{nozaki2009}. As a result, the ionograms carry radio frequency interference (RFI) that changes from frame to frame as vertical bands, together with speckle-like receiver noise (Fig.~\ref{fig:method}, first panel).

Published methods for ionogram cleaning focus on scaling and keep a line-like echo trace. They threshold curvelet coefficients \cite{chen2013}, keep the largest connected region and thin it to a line \cite{yang2017}, repair and thin the F trace \cite{wang2024}, or follow curves with a line detector \cite{fagre2021}. Spread F has been reported as one of their limitations \cite{fagre2021}.

Machine learning models that learn echo structures from ionograms, such as spread-F classifiers \cite{benchawattananon2024}, see noise and echo together unless the images are cleaned first.

We propose a method based on density-based clustering (DBSCAN) that strictly cleans ionogram noise and can handle a variety of echo types including spread F. The method was tuned on 100 hand-picked frames and evaluated on a random set of 100 frames, including its failure cases. The code, grades and examples are shared in the project repository.

\section{Method}
\label{sec:method}
\begin{figure*}[!t]
\centering
\includegraphics[width=\textwidth]{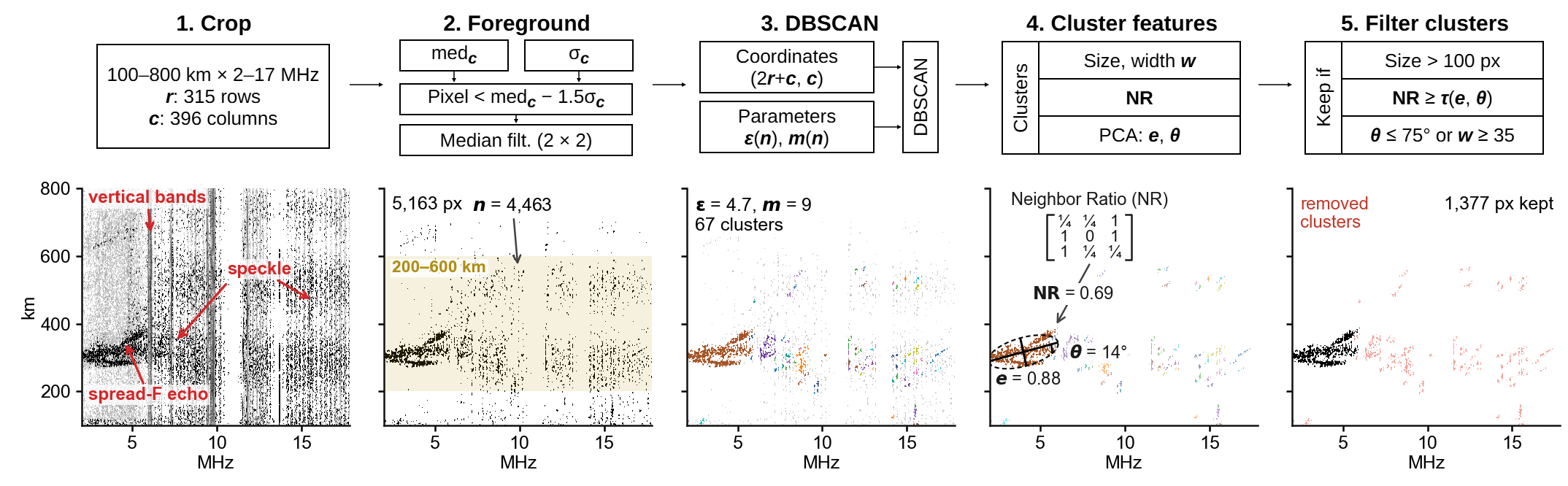}
\caption{Steps of the denoiser on a spread-F frame (CPN, 2025-08-23 15:05 UTC). Panel 3: one color per cluster, unclustered pixels gray. Panel 4: the largest cluster, with the kernel $K$ behind the neighbor ratio NR and the principal axes that give $\theta$ and $e$. Panel 5: kept cluster black, removed clusters red.}
\label{fig:method}
\end{figure*}

The ionograms come from two stations: Chumphon, Thailand (CPN) and Bac Lieu, Vietnam (BCL). Each sounding is an 8-bit gray image of $500 \times 738$ pixels with the echo in black. Fig.~\ref{fig:method} shows the five steps of the proposed denoising process.

\textbf{1. Crop:} The image is cropped to 100 to 800~km and 2 to 17.8~MHz (Fig.~\ref{fig:method}, first panel), which focuses on the F-region echoes.

\textbf{2. Background filter:} A vertical RFI band is nearly constant along height, so the median of each column ($\mathrm{med}_c$) gives the level of the band. A pixel is kept as foreground if it is more than $1.5\,\sigma_c$ darker than $\mathrm{med}_c$, where $\sigma_c$ is the spread of the column with the echo pixels left out. Afterwards, to reduce speckle noise, we apply a $2 \times 2$ median filter, unless the foreground has fewer than 1,500 pixels, in which case the echo is generally already weak and sparse.

\textbf{3. DBSCAN:} The vertical bands are now largely removed, but speckle remains (Fig.~\ref{fig:method}, second panel). It has the same gray level as the echo, so it cannot be separated by intensity. We therefore cluster the foreground with DBSCAN \cite{ester1996} on the coordinates $(2r + c,\, c)$, with $r$ the row from the top and $c$ the column. These coordinates favor rising echo traces and penalize vertical structures: a step along a flat or rising trace has length $\sqrt{2}$, a vertical step 2 and a falling step $\sqrt{10}$. We also set the DBSCAN parameters from $n$, the number of foreground pixels in the 200 to 600~km band. A crowded band means heavy speckle, so clustering is strict, with a small radius $\varepsilon$ and many neighbors $m$. A sparse band gets a large $\varepsilon$ and a small $m$, so a broken trace still forms one cluster.

\textbf{4. Filtering clusters:} Only some of the DBSCAN clusters are echo. To differentiate between echo and noise, we apply a series of rules to each cluster. Each cluster is judged by its size, column width, density and shape. Density is the neighbor ratio (NR), computed for each pixel of a cluster with the kernel $K = [\tfrac{1}{4}, \tfrac{1}{4}, 1;\ 1, 0, 1;\ 1, \tfrac{1}{4}, \tfrac{1}{4}]$, which counts flat and rising neighbors fully and the others at a quarter. Shape comes from a principal component analysis of the pixel positions: $\theta$ is the angle of the main axis from the frequency axis, and $e$ goes from 0.5 for a round cluster to 1 for a line. A cluster is kept if it has more than 100 pixels, if its NR reaches a threshold $\tau$, and if it is not a narrow vertical stripe ($\theta > 75^\circ$, fewer than 35 columns). $\tau$ goes from 0.40 to 0.60 as the frame gets noisier, and thin tilted clusters only need $0.75\,\tau$. On very sparse echo frames the size rule is dropped, and small pieces right next to the kept echo are taken back. The exact parameters of each rule are given in the released code.

A spread-F echo forms one large cluster with a high NR, but heavy speckle can be as dense, so the shape check removes clouds that are neither flat nor dense. In Fig.~\ref{fig:method} the echo passes with an NR of 0.69 against $\tau = 0.60$, while the next largest cluster (NR 0.48, $e = 0.74$) is dropped. The whole pipeline takes 36~ms per frame on average on an Intel Core i7-6800K.

\section{Evaluation}
\label{sec:eval}
Because the 100 tuning frames were picked by hand, the main results are reported on a random set of 100 night frames from the same two stations, 50 of which come from nights with a GNSS scintillation event. Frames were labeled in two ways from the raw crop: by type (echo trace, spread F, or no echo) and, for echo frames, by case (moderate, heavy RFI / speckle, and sparse frame / weak echo).  Heavy RFI / speckle are the frames with either very strong vertical bands overlapping echo or a dense speckle cloud all over the image. Sparse frames have little noise and usually a weak echo. Noise-only frames are common in some periods. An ML model would read any noise left in them as a false echo, so their output should be empty.

With no clean reference to compare against, each output was graded by eye: the echo as kept whole, partly kept (minor gaps are allowed) or lost, and the noise as clean or left. On no-echo frames, clean means an empty output. On the random set (Table~\ref{tab:grading}), the echo was kept whole in 58 of 71 echo frames, partly in 6 and lost in 7, and 87 of 100 outputs were clean. On the tuning set, the same figures were 62 of 82, 18 and 2, and 82 of 100.  

\begin{table}[!t]
\caption{Manual Grading Results on the Random Set\label{tab:grading}}
\centering
\setlength{\tabcolsep}{4pt}
\begin{tabular}{l|c|ccc|cc}
\hline
\multirow{2}{*}{Type / case} & \multirow{2}{*}{$N$} & \multicolumn{3}{c|}{Echo} & \multicolumn{2}{c}{Noise}\\
 & & Kept & Partial & Lost & Clean & Left\\
\hline
\textit{Echo trace} & & & & & & \\
\quad Moderate            & 26 & 24 & 2 & 0 & 23 & 3\\
\quad Heavy RFI / speckle & 17 & 9  & 2 & 6 & 14 & 3\\
\quad Sparse / weak echo  & 7  & 6  & 1 & 0 & 6  & 1\\
\textit{Spread F} & & & & & & \\
\quad Moderate            & 11 & 10 & 1 & 0 & 11 & 0\\
\quad Heavy RFI / speckle & 5  & 4  & 0 & 1 & 4  & 1\\
\quad Sparse / weak echo  & 5  & 5  & 0 & 0 & 5  & 0\\
\textit{No echo}          & 29 & -- & -- & -- & 24 & 5\\
\hline
Total                     & 100 & 58 & 6 & 7 & 87 & 13\\
\hline
\end{tabular}
\end{table}

\begin{figure}[!t]
\centering
\includegraphics[width=0.98\columnwidth]{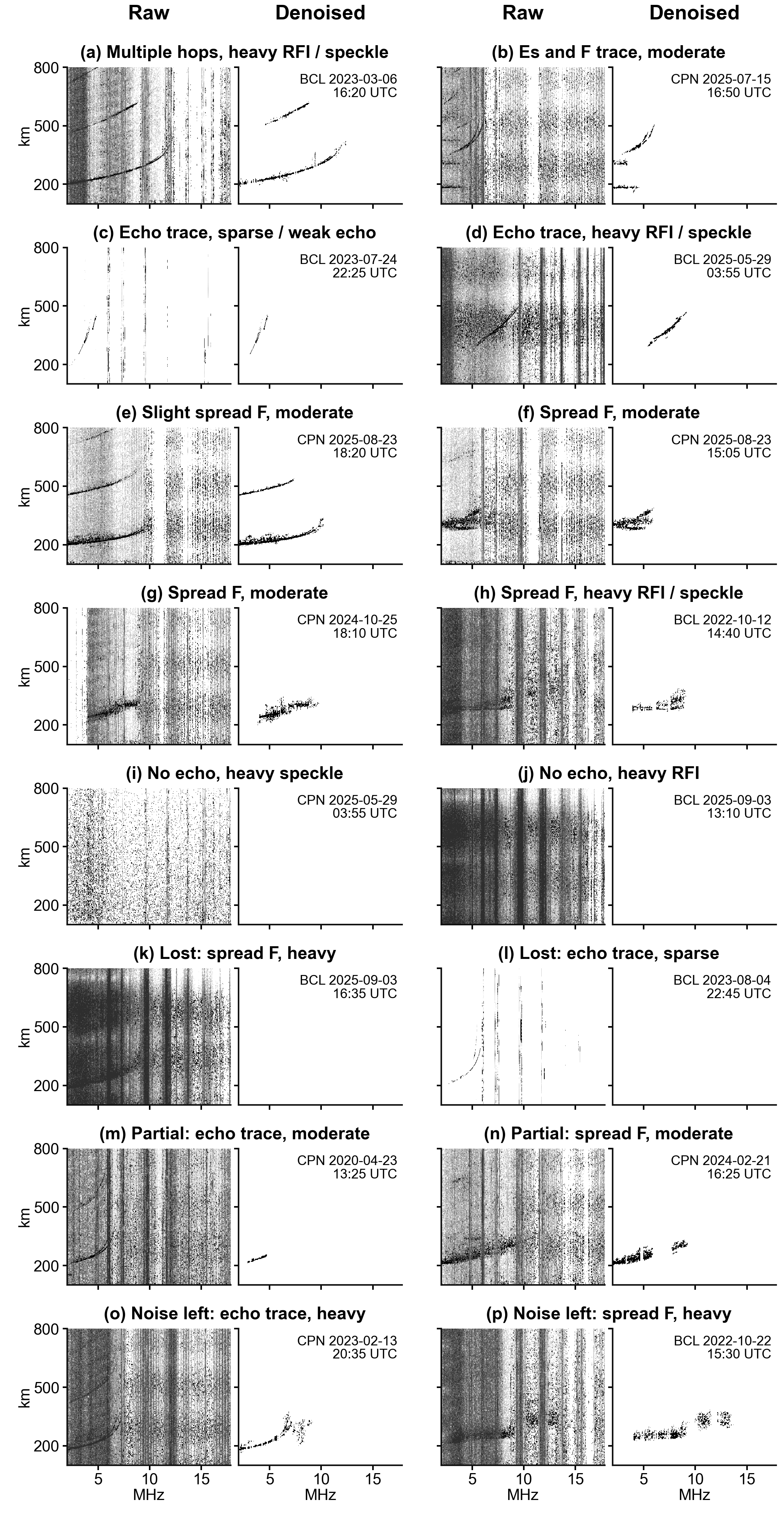}
\caption{The denoiser on 16 graded frames: raw crop (left) and output (right) of each pair, from both the tuning and random sets, named by type, case and grade. Heavy: heavy RFI / speckle. Sparse: sparse / weak echo. (k)--(p) Failures.}
\label{fig:gallery}
\end{figure}

Fig.~\ref{fig:gallery} shows 16 graded frames from both sets, six of them failures. Multiple hops under heavy RFI (a) and sporadic E with moderate noise (b) are kept, as are a weak echo in a sparse frame (c) and an echo under heavy RFI (d). Spread-F echoes from slight to heavy noise pass whole (e)--(h), and noise-only frames come out empty (i), (j).

\section{Limitations and Future Work}
\label{sec:limits}
The proposed denoiser is a heuristic pipeline, tuned by trial and error on CPN and BCL frames. Its pixel rules are tied to 2.2~km and 0.04~MHz per pixel. They must be retuned for another ionosonde or image resolution.

Under moderate noise the denoiser rarely fails. All 7 lost echoes in Table~\ref{tab:grading} are in heavy RFI or speckle frames. The hardest case is an echo buried in a dark band that is not $1.5\,\sigma_c$ darker than its column median, so it is lost at the background filter (k). A faint, disconnected trace in a sparse frame can be lost at the cluster filters (l). Clustering can split a broken trace into pieces, so only part of it is kept (m), (n). Under heavy noise, some speckle can remain next to the echo (o), (p).

Apart from (k), which cannot be separated from the background, these losses are a trade-off. Relaxing the DBSCAN parameters would keep more of the echo but let more noise pass, including in empty frames (i) and (j). We chose to remove noise strictly: some echo is lost in heavy noise, but the output is cleaner and safer for ML tasks that learn physical structure from the echo.

Overall, the denoiser handles a wide variety of echo types and noise conditions, but under heavy noise no fixed set of rules can clean every frame, because the noise changes from frame to frame and across each image. Future work could improve the results in three ways. 1) Echo pieces separated by a gap could be reconnected during clustering, which would reduce lost and partial echoes. 2) The cleaned frames could show whether the noise at each station repeats, for example at fixed frequencies or times of day, so that such a pattern could be removed directly. 3) The noise could be handled in the raw signal rather than in the rendered image, using the full dynamic range of echo and noise. This may be the only way to recover echoes buried under heavy noise.

\section*{Acknowledgments}
This work was supported by Nanyang Technological University (NTU) and Thales Alenia Space. The CPN ionosonde is operated by the National Institute of Information and Communications Technology (NICT), Japan, in collaboration with King Mongkut's Institute of Technology Ladkrabang (KMITL), Thailand. The BCL ionosonde is operated by NICT in collaboration with the Institute of Earth Sciences, Vietnam Academy of Science and Technology (IES-VAST), Vietnam. Claude (Anthropic) was used for programming, language and figure editing support during manuscript preparation. All scientific content, results, and conclusions are the authors' own.

\section*{Data Availability Statement}
SEALION ionograms are available from NICT at \url{https://aer-nc-web.nict.go.jp/sealion/} subject to the SEALION data policy. The denoiser code and the grades are available at \url{https://github.com/Mert-chan/ionogram-denoiser}.

\bibliographystyle{IEEEtran}
\bibliography{ref}

\end{document}